\documentclass[preprint,12pt]{elsarticle}

\journal{Physica A}

\usepackage{graphicx}
\usepackage{color}
\usepackage{dcolumn}
\usepackage{amsmath}
\usepackage{CJK}
\usepackage{subfigure}
\usepackage{sidecap}

\usepackage{mathrsfs}
\usepackage{amsfonts}
\usepackage{indentfirst}
\usepackage{bm}

\newcommand{\be}{\begin{equation}}
\newcommand{\ee}{\end{equation}}
\newcommand{\bey}{\begin{eqnarray}}
\newcommand{\eey}{\end{eqnarray}}
\newcommand{\bw}{\begin{widetext}}
\newcommand{\ew}{\end{widetext}}

\newcommand{\ra}{\rangle}
\newcommand{\la}{\langle}

\newcommand{\ba}{\begin{array}}
\newcommand{\ea}{\end{array}}
\newcommand{\bi}{\begin{itemize}}
\newcommand{\ei}{\end{itemize}}
\newcommand{\bem}{\begin{enumerate}}
\newcommand{\eem}{\end{enumerate}}

\begin{document}

\begin{frontmatter}

\title{Signatures of quantum chaos in the dynamics of bipartite fluctuations}

\author{Qian Wang\corref{cor1}}
\ead{qwang@zjnu.edu.cn}
\cortext[cor1]{Corresponding author.}

\address{Department of Physics, Zhejiang Normal University, Jinhua 321004, China \\
Center for Theoretical Physics of Complex Systems, Institute for Basic Science, Daejeon 34051, Korea\\
CAMTP-Center for Applied Mathematics and Theoretical Physics, University of Maribor, Mladinska 3, SI-2000
Maribor, Slovenia, European Union}


\begin{abstract}
  We study the signatures of quantum chaos by using the concept of bipartite fluctuations in 
  the kicked two-site Bose-Hubbard model, which can be mapped to the well-studied kicked top model. 
  We explore the signatures of quantum chaos in various dynamical properties of bipartite fluctuations.
  The time evolutions of bipartite fluctuations exhibit strongly depends on the locations of 
  initial coherent states in the classical phase space and, therefore, yields the signatures of quantum chaos.
  We further show that the evolution behavior of bipartite fluctuations displays 
  a close relationship to the quantum participation ratio of evolved coherent states on the particle number basis.
  Finally, we discuss how to identify the onset of quantum chaos via the concept of bipartite fluctuations,
  as well as its comparison with the classical Lyapunov exponent.
\end{abstract}



\end{frontmatter}

 \section{Introduction}

 It has been widely known that the classical chaos is characterized by exponential
 separation of trajectories stems from hypersensitivity to initial conditions \cite{ott}.
 However, due to the unitary evolution in quantum mechanics, the above definition of
 chaos can not be translated into the quantum realm.
 Therefore, one of major focus in quantum chaos field is to identify the
 signatures of chaos in quantum systems \cite{stock,haake2}.
 Over the past few decades, several signatures of quantum chaos have been revealed.
 It has been found that in many cases the energy spectral properties 
 \cite{haake2,mcgu,obmj,berry1,berry2,reichl,Atas2013,Chavda2013,Chavda2014,Tekur2018,Sarkar2019,Corps2019}
 is useful to signal the quantum chaos.
 Furthermore, the dynamical signatures of quantum chaos are investigated via the
 Loschmidt echo \cite{peres,schack,agra,Fine2014}, spin squeezing \cite{songd}, 
 and the dynamical generation of the quantum correlations,
 such as entanglement \cite{kfmc,aksh,jnba,madhok,qxie,miller,ghose,piga,xwsg,sgbcs,mam,joshua,kumari,SwSc2019},
 concurrence \cite{xwsg,Dogra2019,dogra}, as well as quantum discord \cite{dogra,gupta}.
 In recent years, the quantum chaos has also been probed by
 the out-of-time ordered correlators \cite{Ivan2017,ebrs,ebrs1,xctz,asvm,gmms,jorge,Fortes2019,Silvia2019}, 
 the quantum Fisher information \cite{Gietka2019}, quantum participation ratio \cite{Miguel2016,Miguel2017},
 and the correlation hole \cite{Torres2017,Torres2018,Torres2019,Hernandez2019}.
 It is worth mentioning that some very recent works have shown that the signatures of quantum chaos, such as the
 exponential growth of the out-of-time ordered correlators, 
 can also appear in classical integrable systems \cite{Hummel2019,Cameo2019}.
 These results indicate that the relationship between quantum and classical chaoses is rather subtle. 
 Therefore, more works are still need to get a better and deeper understanding of the signatures of quantum chaos.

 In this work, we show for the first time that the signatures of quantum chaos can be identified by using
 the concept of bipartite fluctuations \cite{gioev,rachel}.
 In quantum many-body systems, bipartite fluctuations of particle number or magnetization of the subsystem
 has been used as an efficient tool to detect quantum phase transitions for both one-
 and higher-dimension systems \cite{rachel}. In particular, it was founded that in one dimension strongly correlated
 systems, the quantum critical points estimated from bipartite fluctuations with much better accuracy than
 the ones provided by the von Neumann entanglement entropy.
 Moreover, the concept of bipartite fluctuations was also employed to study 
 the topological quantum phase transitions Ref.~\cite{Loic2017}. 
 The connections between bipartite fluctuations and the entanglement lead to
 on the one hand many striking insights into the entanglement properties 
 in quantum many-body systems \cite{hfsong,hfss,flindt,aphf}. 
 On the other hand, the concept of bipartite fluctuations has been established as a means of 
 detecting many-body localization transitions in various systems \cite{luitz,baygan,singh,lee,Singh2017}.
 In the Heisenberg model, it has been found that bipartite fluctuations of the magnetization increases 
 with the system size and reaches a constant in ergodic phase, 
 while it becomes vanishingly small and independent of system size in many-body localization phase \cite{singh,lee}.
 One of a particular advantage of bipartite fluctuations is that it can be measured in
 experiments by using, for example, single atom microscopy \cite{bakr,sherson}.
 Thus, the signatures of quantum chaos obtained from bipartite fluctuations are particularly valuable
 for an experimental study of the quantum chaos.
 The other advantage of bipartite fluctuations is that it is easier to calculate
 numerically than the quantities, such as the out-of-time ordered correlators 
 and entanglement entropy, in many-body systems.

 The aim of the present work is to explore how the signatures of quantum chaos
 manifest themselves in the dynamical behaviors of bipartite fluctuations.
 To this end, we study the dynamical properties of bipartite fluctuations in
 the kicked two-site Bose-Hubbard (BH) model \cite{qxie,mps,ckdc},
 which can be mapped to the kicked top model via Schwinger representation.
 The kicked top model is a standard model for studies of quantum chaos and
 has rich dynamical features in both classical and quantum regimes \cite{haake2}.
 By varying the locations of the initial coherent state from the regular to chaotic
 regions in the classical phase space, we will identify the signatures of quantum chaos 
 through the time evolutions of bipartite fluctuations, as well as the long-time averaged bipartite fluctuations.
 We also demonstrate that the evolution of bipartite fluctuations is closely related to the localization property of the
 evolved coherent state on the particle number basis. 
 We further show how to characterize the regular-chaotic transition in quantum system 
 via the concept of bipartite fluctuations.

 The rest of this article is organized as follows.
 In Sec.~\ref{model}, we introduce the kicked two-site BH model and its basic features.
 We also discuss its classical counterpart in this section.
 In Sec.~\ref{results}, we analysis in detail the dynamics of bipartite fluctuations.
 We show that both the local and global signatures of quantum chaos can be revealed via the  dynamical properties
 of bipartite fluctuations.
 Finally, we give our conclusion and discuss our results in Sec.~\ref{summary}.

 \section{The model}  \label{model}

 We consider the kicked two-site BH model in which we periodically vary
 the hopping term by a sequence of kickings, the Hamiltonian is, therefore, given by \cite{qxie,mps}
 \be \label{KBHH}
    \hat{H}(t)=\frac{U}{4}(\hat{a}_1^\dag\hat{a}_1-\hat{a}_2^\dag\hat{a}_2)^2+
            \frac{V}{2}(\hat{a}_1^\dag\hat{a}_2+\hat{a}_2^\dag\hat{a}_1)\sum_{m=-\infty}^{+\infty}
             \delta(t-m),
 \ee
 where $\hat{a}_j (\hat{a}_j^\dag)$ is the bosonic annihilation (creation) operator for the $j$th site,
 $U$ denotes the strength of the on-site interaction, $V$ determines the tunneling strength.
 The particle number operator $\hat{N}=\hat{n}_1+\hat{n}_2$ with $\hat{n}_1=\hat{a}_1^\dag\hat{a}_1
 (\hat{n}_2=\hat{a}_2^\dag\hat{a}_2)$, commutates with
 the Hamiltonian, thus the total number of particles $N$ is a conserved quantity.

 \begin{figure}
  \includegraphics[width=\columnwidth]{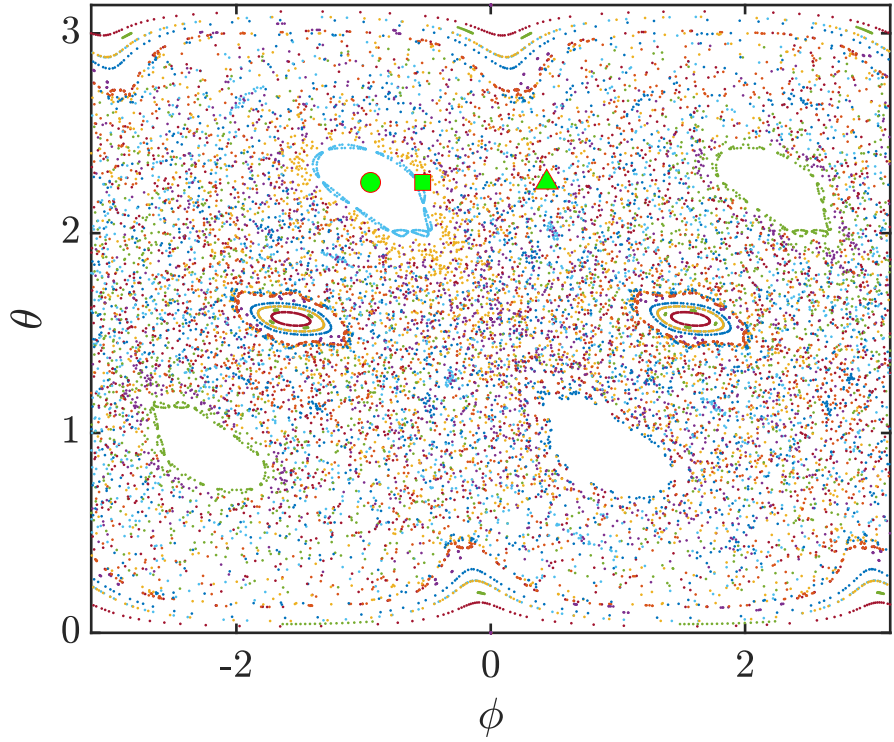}
  \caption{Classical phase space of the kicked two-site BH model (kicked top model) with $\kappa=3$. The classical
   variables ($\theta, \phi$) are plotted after each kick for $157$ trajectories, each with a duration of $300$ kicks.
   The filled green circle marks the fix point of the regular region with ($\theta,\phi$)=$(2.254,-0.945)$.
   The filled green square indicates the
   boundary between the regular and chaotic regions with ($\theta,\phi$)=$(2.254,-0.535)$.
   The filled green triangle at ($\theta,\phi$)=$(2.254,0.44)$ is in the chaotic sea.}
  \label{CPC}
 \end{figure}

 By introducing the angular momentum operators via Schwinger representation \cite{mps}:
 $\hat{J}_x=(\hat{a}_1^\dag\hat{a}_2+\hat{a}_2^\dag\hat{a}_1)/2$, 
 $\hat{J}_y=(\hat{a}_1^\dag\hat{a}_2-\hat{a}_2^\dag\hat{a}_1)/(2i)$,
 and $\hat{J}_z=(\hat{a}_1^\dag\hat{a}_1-\hat{a}_2^\dag\hat{a}_2)/2$.
 The Hamiltonian in Eq.~(\ref{KBHH}) can be mapped to the well-known quantum kicked top model \cite{haake2,mps,ckdc,haake1}
 \be \label{QKT}
    \hat{\mathcal{H}}(t)=\frac{\kappa}{2j}\hat{J}_z^2+V\hat{J}_x\sum_{m=-\infty}^{+\infty}\delta(t-m),
 \ee
 where $\kappa=NU$ and $j=N/2$ is a conserved quantity.
 Since $N$ is conserved, $j$ is, therefore, a constant of motion.
 The dimension of the Hilbert space of the quantum kicked top model is $\mathcal{D}_\mathcal{H}=2j+1$.
 Therefore, on can explore the dynamics of the quantum kicked model without truncating the Hilbert space.
 The quantum kicked top is a standard paradigm for both theoretical 
 \cite{ghose,piga,xwsg,sgbcs,mam,joshua,kumari,dogra,gupta,ckdc,haake1,fox,mksg,utbm} 
 and experimental \cite{scas,neill} studies of quantum chaos.
 Its classical counterpart has a range of dynamics from regularity to fully chaotic.
 Without loss of generality, in the following of our study we set $V=\pi/2$.
 The results for other values of $V$ are qualitatively similar.

 Under the Hamiltonian (\ref{QKT}), the system state at time $t$ reads $|\psi(t)\ra=\hat{\mathcal{U}}|\psi(0)\ra$,
 where $|\psi(0)\ra$ is the initial state and $\hat{\mathcal{U}}$ denotes the unitary evolution operator, i.e., 
 $\hat{\mathcal{U}}=\hat{\mathcal{T}}\exp[-i\int_0^t\hat{\mathcal{H}}(t)dt]$ with $\hat{\mathcal{T}}$ 
 is the time-ordering operator.
 For the kicked system, $\hat{\mathcal{U}}$ is given by the Floquet operator \cite{stock,haake2}
 \be
    \hat{\mathcal{U}}=\exp\left(-i\frac{\pi}{2}\hat{J}_x\right)\exp\left(-i\frac{\kappa}{2j}\hat{J}_z^2\right).
 \ee
 Then, after $n$th kick, the state of the system evolves to $|\psi(n)\ra=\hat{\mathcal{U}}^n|\psi(0)\ra$.

 In the classical limit, i.e., $N\to\infty$ ($j\to\infty$), by using the Heisenberg equation of the angular momentum operators,
 the classical mapping between the consecutive kicks can be written as \cite{qxie,piga,fox}
 \begin{align}
    X_{n+1}&=X_n\cos(\kappa Y_n)+Z_n\sin(\kappa Y_n), \notag \\
    Y_{n+1}&=X_n\sin(\kappa Y_n)-Z_n\cos(\kappa Y_n),  \label{ClassicalM}\\
    Z_{n+1}&=Y_n, \notag
 \end{align}
 where $X=\la\hat{J}_x\ra/j, Y=\la\hat{J}_y\ra/j$ and $Z=\la\hat{J}_z\ra/j$ are the rescaled angular momentums.
 The conservation of $N$ leads to the constraint $X^2+Y^2+Z^2=1$, which means that
 the classical dynamics restricts on the surface of the unit sphere with $(X,Y,Z)=(\sin\theta\cos\phi,
 \sin\theta\sin\phi,\cos\theta)$, where $\theta, \phi$ are the usual polar and azimuthal angles, respectively.
 Hence, the classical phase-space is essentially two dimensional \cite{qxie,miller,xwsg,joshua}.

 The classical dynamics determined by Eq.~(\ref{ClassicalM}) depends on the
 parameter $\kappa$, the so-called chaoticity parameter.
 It has been already known that the classical trajectories
 are regular for small values of $\kappa$, whereas, the dynamics of the system
 crosseover to fully chaotic motion with increasing $\kappa$ \cite{miller,ghose,mps,mksg}.
 In Fig.~\ref{CPC}, we plot the classical phase space of the kicked top with $\kappa=3$.
 Obviously, the phase space shows a mixed feature, the regular islands are embedded in the chaotic sea.
 In the following of our study, we will focus on three different initial conditions, 
 as is shown in Fig.~\ref{CPC} (see the caption of Fig.~\ref{CPC} for details).
 As we will see,  the signatures of quantum chaos are highly correlated to the properties of the classical
 phase space.

 \begin{figure}
   \begin{minipage}{0.5\linewidth}
   \centering
   \includegraphics[width=\columnwidth]{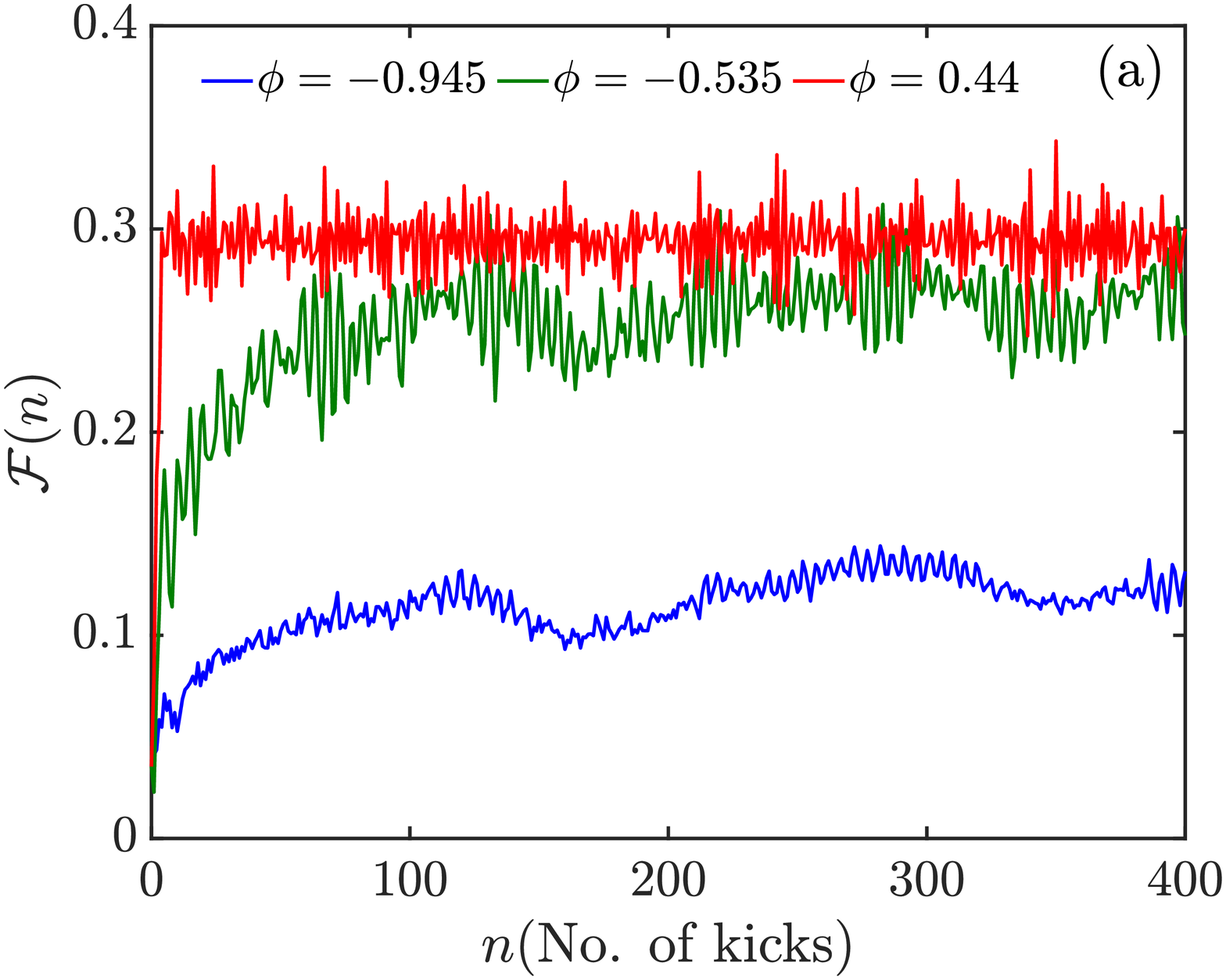}
   \end{minipage}
   \begin{minipage}{0.5\linewidth}
   \centering
   \includegraphics[width=\columnwidth]{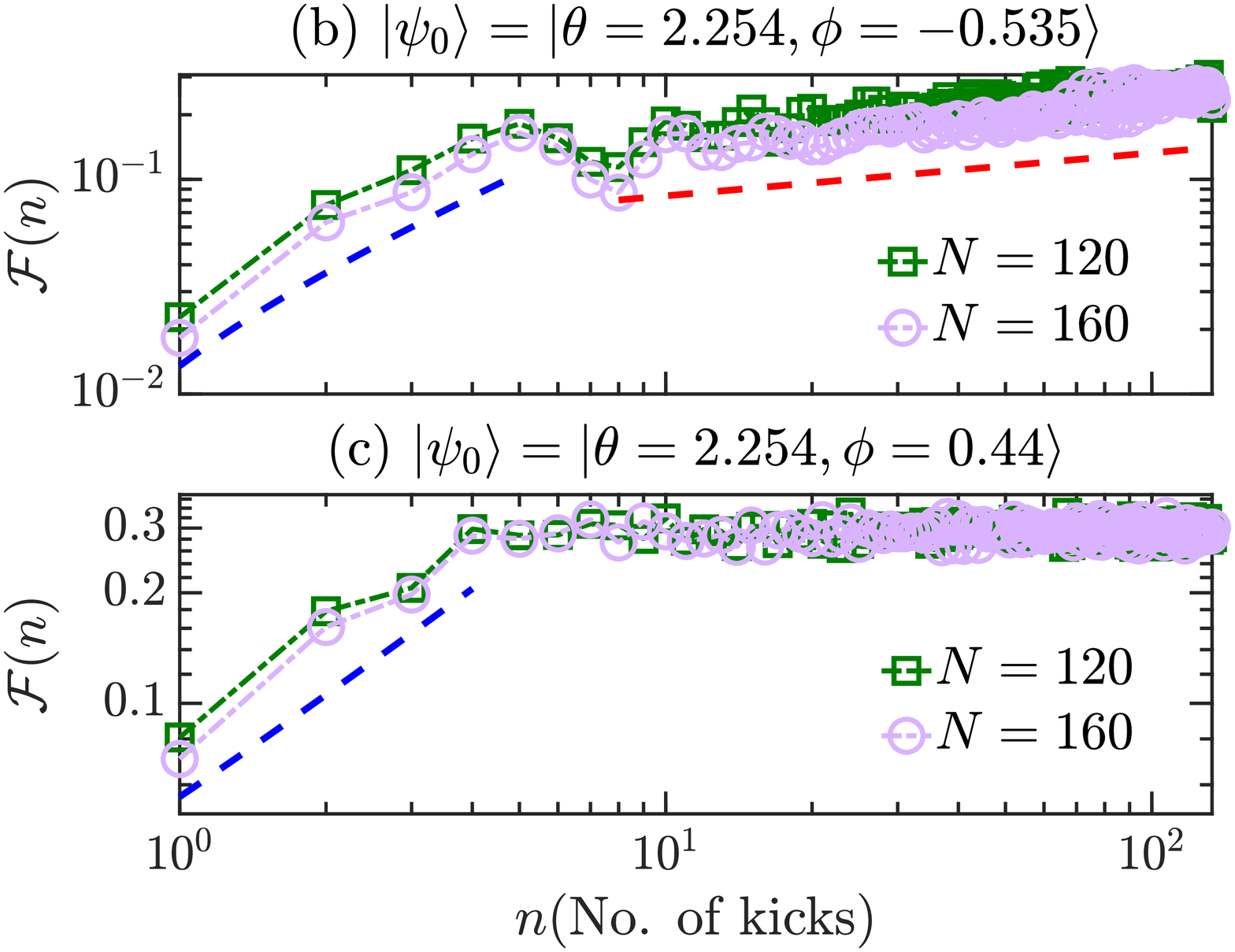} 
   \end{minipage}
   \caption{(a) The time evolution of $\mathcal{F}(n)$ for different
   initial states $|\theta,\phi\ra$ with $\theta=2.254$, $N=120$.
   (b) Log-log plot for the short time behavior of $\mathcal{F}(n)$ for different system size 
   $N$ with initial state $|\theta=2.254,\phi=-0.535\ra$.
   (c) Log-log plot for the short time evolution of $\mathcal{F}(n)$ for different system size $N$ 
   with initial state $|\theta=2.254,\phi=0.44\ra$.  
   The value of the chaoticity parameter for all figures is $\kappa=3$. 
   The blue dashed lines in (b) and (c) denote the linear behavior $\mathcal{F}(n)\propto n$.
   The red dashed line in (b) indicates the polynomial growth $\mathcal{F}(n)\propto n^{0.2}$.}
   \label{timeCBF}
 \end{figure}

 \section{The dynamics of bipartite fluctuations and chaos}  \label{results}

 We now study the dynamics of bipartite fluctuations \cite{rachel}
 in the kicked two-site BH model (kicked top model).
 The subsystems of our model are provided by two sites.
 In spite of the total number of particles is conserved,
 the number of particles in each subsystem fluctuates.
 Consider the rescaled number of particles on the first site $\hat{n}_1/N=\hat{a}_1^\dag\hat{a}_1/N$,
 we define the bipartite fluctuations after $n$th kick as the quantum fluctuations of $\hat{n}_1/N$
 \be
    \mathcal{F}(n)=[\la\psi(n)|(\hat{n}_1/N)^2|\psi(n)\ra-\la\psi(n)|\hat{n}_1/N|\psi(n)\ra^2]^{1/2},
 \ee
 where $|\psi(n)\ra=\hat{\mathcal{U}}^n|\psi_0\ra$ and
 $|\psi_0\ra$ is the initial state of the system.
 In angular momentum representation, we have $\hat{n}_1/N=1/2+\hat{J}_z/N$.
 Note that $\mathcal{F}(n)$ defined here is the square root of bipartite
 fluctuations defined in Ref.~\cite{rachel}. 
  
 In order to reveal the signatures of quantum chaos and compare with the classical chaos, we take the system 
 initially in an arbitrary coherent state. In the Fock basis, the initial coherent state reads
 \cite{mps,ftae,perelomov,gilmore}
 \begin{align} \label{SCS}
  |\psi_0\ra=|\theta,\phi\ra
             =\sum_{l=0}^N\sqrt{\binom{N}{l}}\cos^l\left(\frac{\theta}{2}\right)
              \sin^{N-l}\left(\frac{\theta}{2}\right)e^{i(N-l)\phi}|l\ra,
 \end{align}
 where $|l\ra=|l,N-l\ra=(\hat{a}_1^\dag)^l(\hat{a}_2^\dag)^{N-l}|0,0\ra/\sqrt{l!(N-l)!}$
 with $l=0,1,\ldots,N$, is the $l$th eigenstate of $\hat{n}_1$, and $0\leq\theta\leq\pi$, $-\pi\leq\phi\leq\pi$.

 \begin{figure} 
   \begin{minipage}{0.5\linewidth}
   \centering
   \includegraphics[width=\columnwidth]{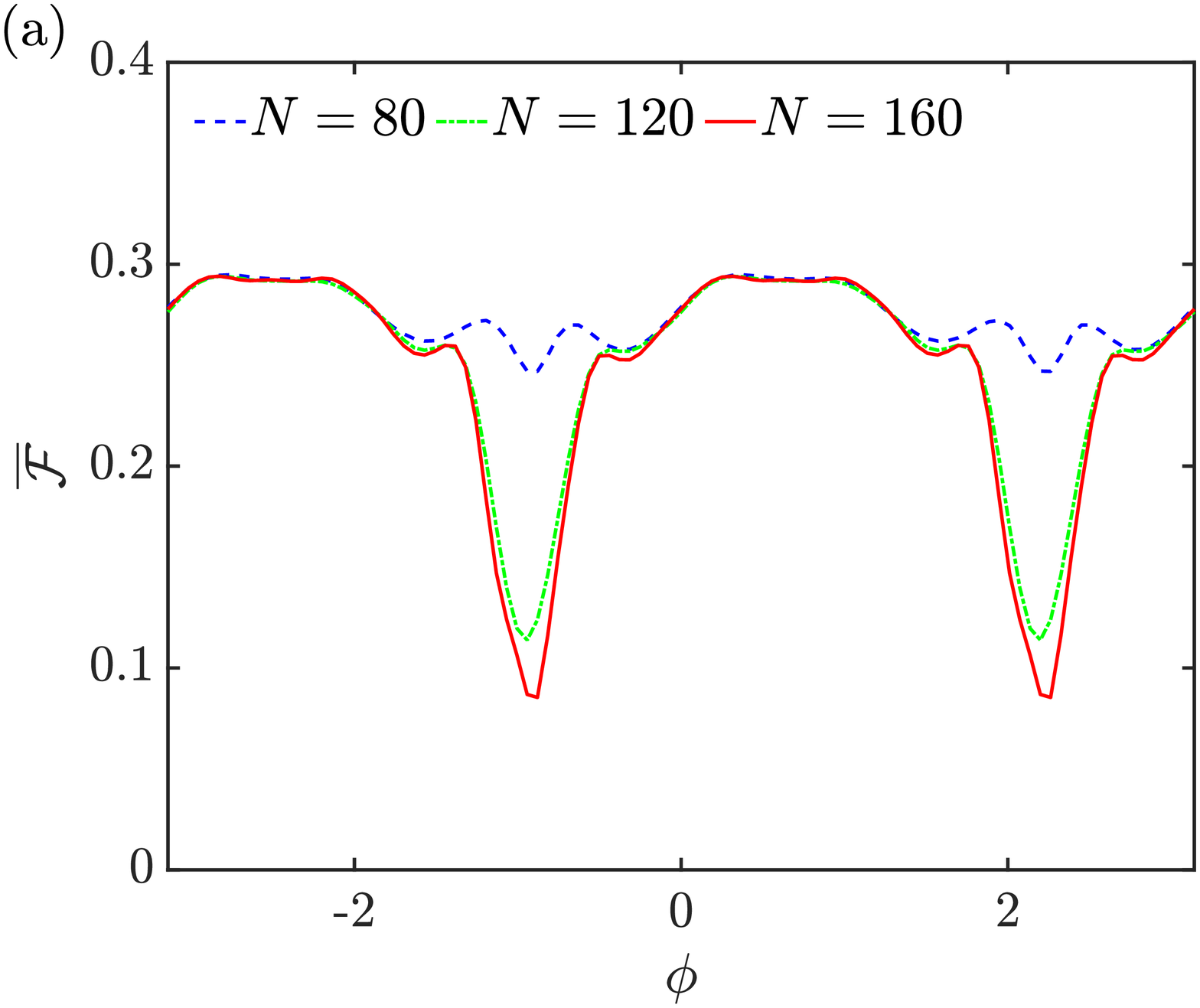}
   \end{minipage} 
   \begin{minipage}{0.5\linewidth}
   \centering
  \includegraphics[width=\columnwidth]{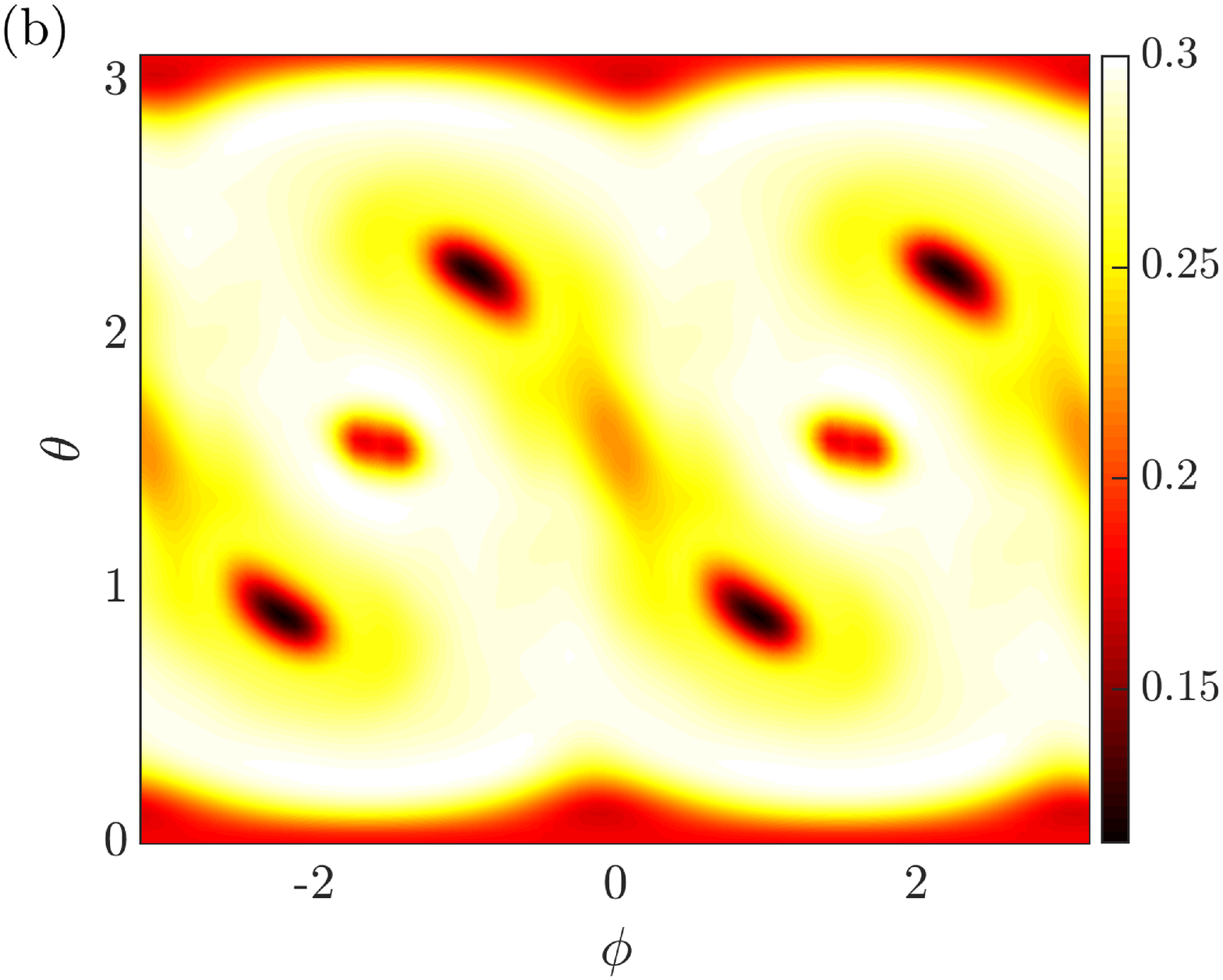}
  \end{minipage}
  \caption{(a) The time averaged bipartite fluctuations as a function of the azimuthal angle $\phi$ with
   $\theta=2.254$ for different system size $N$.
   (b) The time averaged bipartite fluctuations as a function of polar angle $\theta$ and azimuthal angle $\phi$ with $N=160$.
   For both figures, we take $\kappa=3$.}
  \label{PCQ}
 \end{figure}

 We start by investigating the dynamics of $\mathcal{F}(t)$ for different initial states with
 the chaoticity parameter is fixed at $\kappa=3$.
 The initial coherent states are chosen from three different regions of the phase space, namely
 the fixed point of the regular region, the border between the regular and chaotic regions, and the fully
 chaotic region (see the caption of Fig.~\ref{CPC} for details).

 Fig.~\ref{timeCBF}(a) shows the evolution of $\mathcal{F}(n)$
 for aforementioned initial coherent states with $N=120$.
 Several remarkable features can be found from this figure.
 Firstly, since the initial coherent states are localized, the value of $\mathcal{F}(n)$ is small at the initial time.
 As time increases, the initially localized states are starting to spread over time on phase space, which results
 in $\mathcal{F}(n)$ growth with time.
 For the state initialized in the fixed point, $\mathcal{F}(n)$ increases with small amplitude
 and shows a quasiperiodic behavior at late time due to the underlying classical regular dynamics.
 When the initial state centered in the chaotic sea, $\mathcal{F}(n)$ exhibits a fast growth for a very short duration.
 Then it is followed by the random oscillations around its saturation value. 
 The initially rapid growth is due to the fast spreading of the initial state in phase space before 
 saturating because of the finite size effect.
 The intermediate behavior of $\mathcal{F}(n)$ can be observed for the curve with $\phi=-0.535$.
 In this case, after the initial rapid growth, $\mathcal{F}(n)$ shows slow growth for a long time
 before it eventually oscillates randomly about its saturation value.
 Secondly, the quasiperiodic modulation of $\mathcal{F}(n)$ is disappeared as the initial state
 is changed from regular to chaotic regions in classical phase space.
 Finally, we see that the saturates value of $\mathcal{F}(n)$
 is enhanced by the underlying classical chaotic dynamics.
 
 To obtain further insight into the details of the dynamics of $\mathcal{F}(n)$, 
 we present the log-log plot of the evolutions of $\mathcal{F}(n)$ in Figs.~\ref{timeCBF}(b) and (c) for 
 different initial coherent states with several system sizes. 
 In very short time, the evolution of $\mathcal{F}(n)$ for the intermediate state is qualitatively similar to the chaotic case. 
 In both cases, we find $\mathcal{F}(n)$ shows a linear growth with time, irregardless of the system size.
 However, $\mathcal{F}(n)$ displays very different behaviors in the following evolutions.
 Specifically, when the initial state located at the boundary between the regular and chaotic regions, the initial
 linear growth of $\mathcal{F}(n)$ is followed by a slower polynomial growth before it saturates to a finite value. 
 We find that the polynomial growth can be fitted by $\mathcal{F}(n)\propto n^{0.2}$ [cf.~Fig.~\ref{timeCBF}(b)].
 In contrast, there is no polynomial growth for chaotic state and $\mathcal{F}(n)$ saturates to its saturation value
 that independent of the system size [cf.~Fig.~\ref{timeCBF}(c)].
 We note that the duration of polynomial growth increases with increasing the system size.
 The polynomial growth of observables has been investigated in detail 
 in the so-called quantum weakly chaotic systems \cite{Fine2014,Ivan2017,Silvia2019} 
 and diffusive quantum many-body system \cite{Bohrdt2017}.
 It is known the echo and/or OTOC of local observables or sums of them will grow polynomially in time for the systems 
 that lack the well-defined classical limit \cite{Silvia2019}.
 In our study, however, we find that the polynomial growth behavior also existed 
 in the system that has a well-defined classical limit. 

 The general dynamical behaviors of $\mathcal{F}(n)$ in regular and chaotic regions are
 represented by aforementioned three initial states.
 The results display in Fig.~\ref{timeCBF} indicate that the signatures of quantum chaos
 are revealed in the dynamics of bipartite fluctuations.
 Fig.~\ref{timeCBF}(a) further shows that chaos in quantum system is strongly correlated to 
 the saturation value of $\mathcal{F}(n)$, which is defined as the long-time average of $\mathcal{F}(n)$
 \be \label{TAVF}
     \overline{\mathcal{F}}=\lim_{\mathcal{N}\to\infty}\frac{1}{\mathcal{N}}\int_0^\mathcal{N}\mathcal{F}(n)dn,
 \ee
 where $\mathcal{N}$ is the total number of kicks.
 To numerically calculate $\overline{\mathcal{F}}$, $\mathcal{N}$ should be chosen to be larger
 than other time scales.
 In our study, we take $\mathcal{N}=400$, we have checked that the results do not change for larger $\mathcal{N}$.
 The results in Fig.~\ref{timeCBF}(a) allow us to conclude that
 the regular and chaotic regions in phase space can be detected through $\overline{\mathcal{F}}$. 

 \begin{figure}
   \begin{minipage}{0.5\linewidth}
   \centering
   \includegraphics[width=\columnwidth]{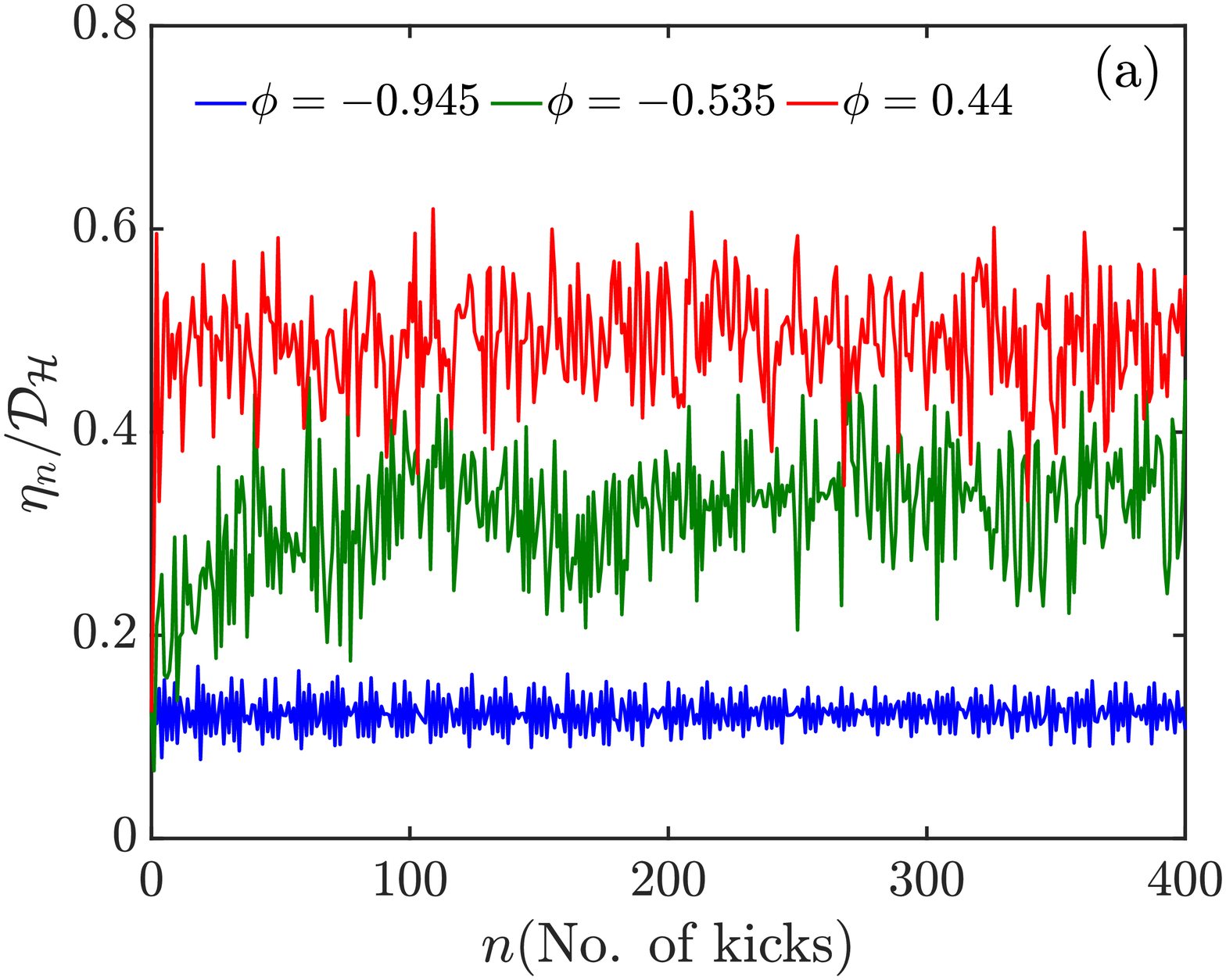}
   \end{minipage}
   \begin{minipage}{0.5\linewidth}
   \centering
   \includegraphics[width=\columnwidth]{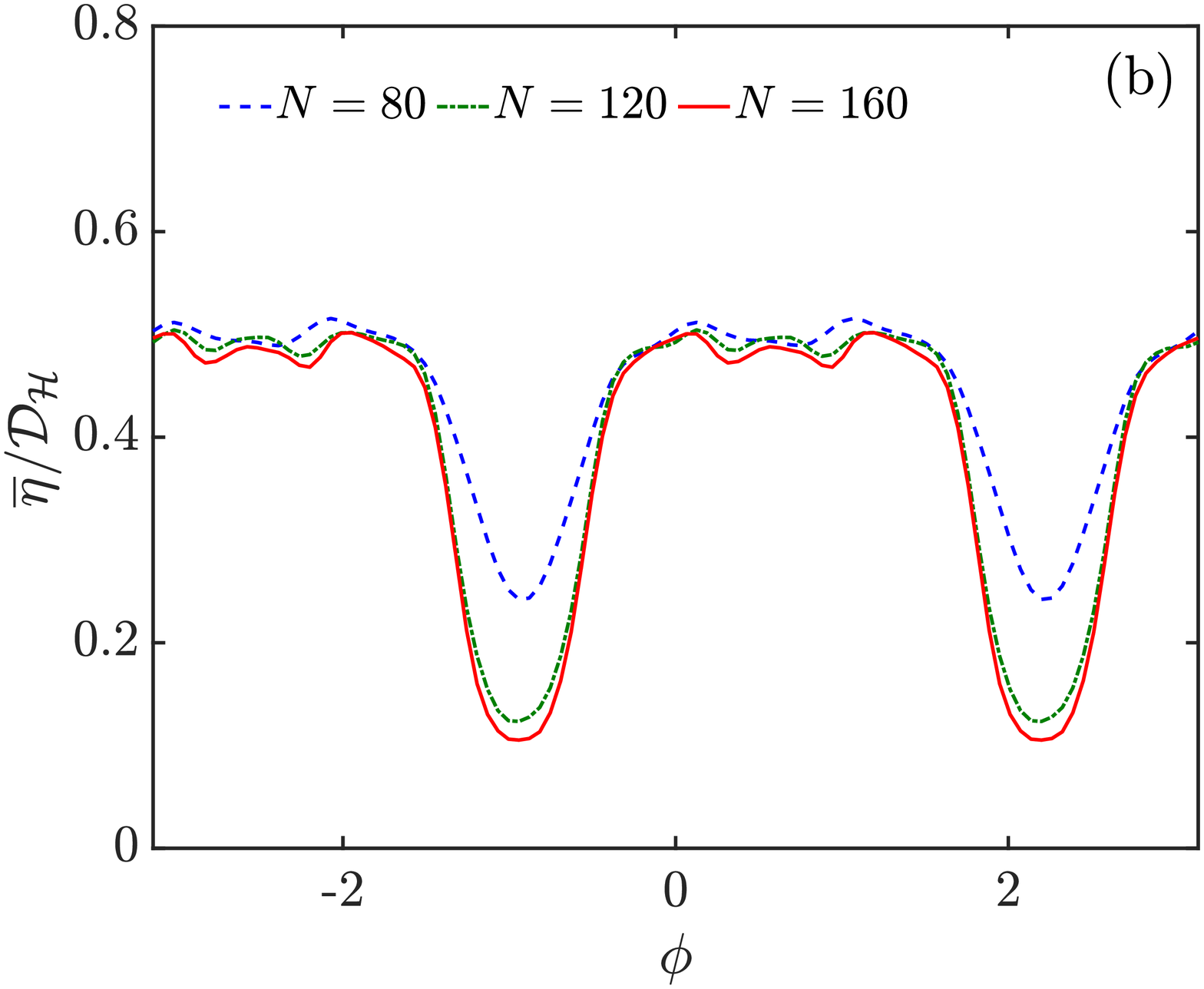}
   \end{minipage}
   \caption{(a) The evolution of the PR for different
   initial coherent states $|\theta,\phi\ra$ with $\theta=2.254$, $N=120$ and $\kappa=3$.
   (b) The long-time averaged PR as a function of the azimuthal angle $\phi$ with
   $\theta=2.254$ for different system size $N$.
   The value of chaoticity parameter is $\kappa=3$ and the time
   average is over $400$ kicks.
   Here both PR $\eta_n$ and the long-time averaged IPR $\bar{\eta}$ are
   rescaled by the dimension of Hilbert space $\mathcal{D}_\mathcal{H}=N+1$.}
   \label{timeIPR}
 \end{figure}

 In Fig.~\ref{PCQ}(a), we plot $\overline{\mathcal{F}}$ as function of
 the azimuthal angle $\phi$ with a constant polar angle $\theta=2.254$ for different system sizes.
 Clearly, the regular regions are identified by the dips in $\overline{\mathcal{F}}$ and the minimum values in 
 $\overline{\mathcal{F}}$ are approximately corresponded to fixed points in phase space.
 The minimum value in $\overline{\mathcal{F}}$ decreases with increasing $N$, while the larger the system size $N$
 the wider dips in $\overline{\mathcal{F}}$.
 As $N$ goes to infinity, we expect that the the regular regions revealed in $\overline{\mathcal{F}}$ 
 will coincide with the corresponding classical ones in Fig.~\ref{CPC}.
 For chaotic regions, in contrast, $\overline{\mathcal{F}}$ saturates to a 
 larger value which independent of the system size $N$.
 Thus, $\overline{\mathcal{F}}$ provides a quantum tool to identify the regular and
 chaotic regions in classical phase space.

 Fig.~\ref{PCQ}(b) displays $\overline{\mathcal{F}}$ as a function of $\phi$ and $\theta$.
 Comparing with Fig.~\ref{CPC}, we clearly see the mixed nature of the classical phase
 space is well reproduced in Fig.~\ref{PCQ}(b).
 In particular, the four evident islands in Fig.~\ref{PCQ}(b) are closely matched with the four stable
 islands in Fig.~\ref{CPC}.
 Therefore, using $\overline{\mathcal{F}}$, we get a good quantum-classical correspondence.

 The features of bipartite fluctuations discussed above are induced by the fact that
 in the regular region, the evolved state of system is a localized state in
 the basis provided by the eigenstate of $\hat{n}_1$,
 whereas in the chaotic region it becomes an extended state.
 To clarify this statement, we study the participation ration (PR) \cite{kota,santos},
 which quantifies the degree of delocalization of the evolved state in the eigenstate of $\hat{n}_1$.
 After $n$th kick, we decompose the evolved state $|\psi_n\ra$ in the basis $|l\ra$,
 $|\psi_n\ra=\sum_l c_l^n|l\ra$, where $|l\ra$ is the $l$th eigenstate of $\hat{n}_1$ [cf.~Eq.~(\ref{SCS})],
 $c_l^n=\la l|\psi_n\ra$ is the expansion coefficient and satisfy $\sum_l|c_l^n|^2=1$.
 Then the PR is defined as
 \be
   \eta_n=\frac{1}{\sum_l |c_l^n|^4}.
 \ee
 The extended state in the basis $|l\ra$ is indicated by the large value of $\eta_n$, while
 small value of $\eta_n$ implies localized state.
 It is worth pointing out that the participation ratio of the coherent states on a suitable basis
 has been proved equivalent to  the classical Lyapunov exponent in the studies of 
 the quantum chaos \cite{Miguel2016,Miguel2017}.

 Fig.~\ref{timeIPR}(a) shows the PR as a function of time for aforemetioned initial coherent states.
 We see that the PR is small for the initial state centered in the fixed point, whereas
 it increases with the initial state moving to the chaotic region.
 This result confirms that in the regular region, the initial coherent state persists its localization nature, whereas 
 it evolves into an extended state in the chaotic region.

 To further elucidate the localization property of the evolved state in regular versus chaotic regions in
 classical phase space, we calculate the rescaled long-time averaged PR $\bar{\eta}/\mathcal{D}_\mathcal{H}$,
 where $\mathcal{D}_\mathcal{H}$ is the dimension of quantum Hilbert space.
 Fig.~\ref{timeIPR}(b) shows $\bar{\eta}/\mathcal{D}_\mathcal{H}$ as a function of
 azimuthal angle $\phi$ with $\theta=2.254$ for different size of system.
 Obviously, $\bar{\eta}/\mathcal{D}_\mathcal{H}$ has significant small values in the regular regions,
 which is consistent with the fact that in the regular regions the evolved state
 remains localized in eigenstates of $\hat{n}_1$.
 In the chaotic region, however, due to the delocalization of the evolved
 state, $\bar{\eta}/\mathcal{D}_\mathcal{H}$ takes a larger value and does not change much with $\phi$.
 Moreover, the value of $\bar{\eta}/\mathcal{D}_\mathcal{H}$ decreases with increasing
 the system size in the regular regions, while 
 it is almost independent of the system size in the chaotic regions.
 The extension of the evolved state on
 the eigenstates of $\hat{n}_1$ results in the bipartite fluctuations grows faster with time
 and quickly saturates to its saturation value.
 By contrast,  in the regular regions, the localization of the evolved state leads to
 the quasiperiodic behavior in the evolution of bipartite fluctuations.

 So far, we have definitely demonstrated that the local signatures of chaos can identify through
 different properties of bipartite fluctuations.
 However, to confirm the concept of bipartite fluctuations is a valuable diagnostic tool in the studies of quantum chaos,
 we also need to know whether it can detect the transition from integrability to chaos.
 In the following, we focus on the phase space averaged $\overline{\mathcal{F}}$, which we call as the bipartite
 fluctuations power and denote by $\la \overline{\mathcal{F}}\ra_\mathcal{P}$, i.e.,
  \be
   \la\overline{\mathcal{F}}\ra_\mathcal{P}=\int d\mu(\theta,\phi)\overline{\mathcal{F}}.
 \ee
 Here, $\overline{\mathcal{F}}$ is given by Eq.~(\ref{TAVF}) and
 $d\mu(\theta,\phi)=\sin\theta d\theta d\phi$ \cite{miller}
 is the Haar measure of the phase space. We will investigate the behavior of $\la\overline{\mathcal{F}}\ra_\mathcal{P}$ 
 as a function of the chaoticity parameter $\kappa$.
 
 As is well known, in classical systems, the degree of chaos is measured by the Lyapunov exponent (LE).
 It equals zero in the regular regimes, while in the chaotic regimes it takes
 non-zero positive value and increases with the degree of chaos in the system.
 For quantum systems,
 the participation ratio of the coherent state has been established as a means of diagnosing quantum chaos 
 and, in particular, plays a role as the classical LE in quantum chaotic systems \cite{Miguel2016,Miguel2017}.
 A close correlation between the evolution behaviors of bipartite fluctuations and the quantum participation ration 
 of evolved coherent states on the particle number basis allows us to expect that, 
 such as the  classical LE, the bipartite fluctuations power can also identify the transition from integrability to chaos.

 \begin{figure}
  \includegraphics[width=\columnwidth]{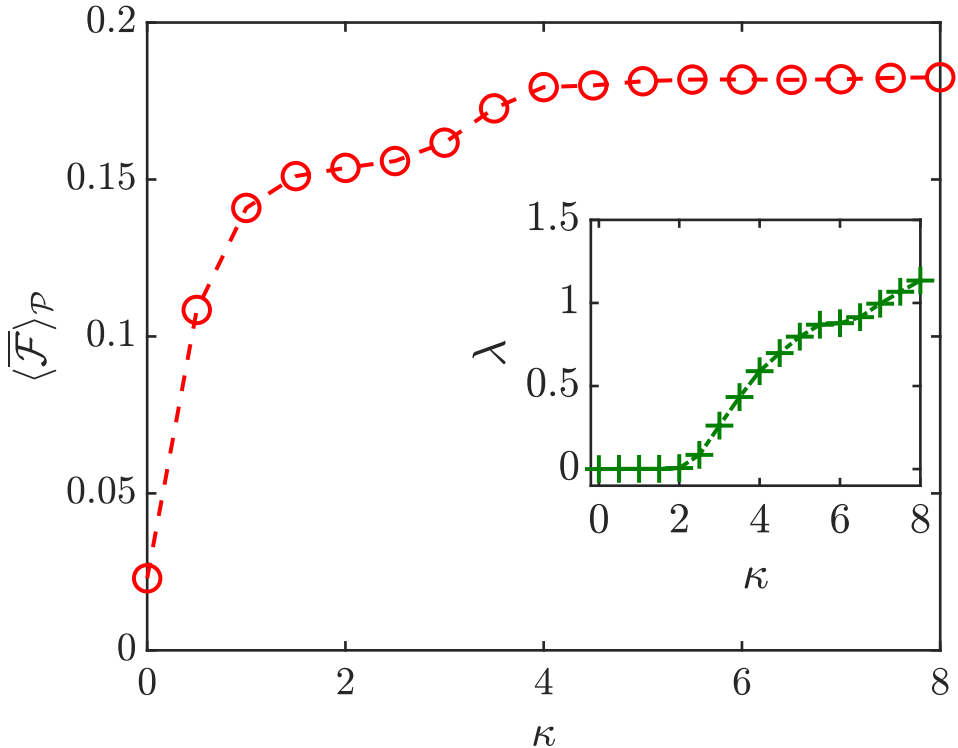}
  \caption{$\la\overline{\mathcal{F}}\ra_\mathcal{P}$ as a function of $\kappa$ for $N=160$.
   The phase space average is calculated on a grid of $15\times15$ initial conditions
   and the time average is taken over $400$ kicks.
   Inset: The Lyapunov exponent $\lambda$ versus the chaoticity parameter $\kappa$.
   }
  \label{PFLY}
 \end{figure}

 In Fig.~\ref{PFLY}, we plot $\la\overline{\mathcal{F}}\ra_\mathcal{P}$
 as a function of chaoticity parameter $\kappa$.
 To confirm $\la\overline{\mathcal{F}}\ra_\mathcal{P}$ is indeed able to detect the onset of quantum chaos and to study the
 classical-quantum correspondence, we also plot the classical LE, which we denote as $\lambda$, in the same figure
 (see the inset of Fig.~\ref{PFLY}).
 Here, it should be noted out that in our study the classical LE is numerically calculated
 by using the tangent map of Eq.~(\ref{ClassicalM}) (see Ref.~\cite{piga} for details).
 
 As is shown in Fig.~\ref{PFLY}, for small $\kappa$, $\la\overline{\mathcal{F}}\ra_\mathcal{P}$ exhibits a fast growth which
 then followed by a slow growth for $\kappa$ ranges approximately from $1.8$ to $2.2$.
 The slow growth of $\la\overline{\mathcal{F}}\ra_\mathcal{P}$ is terminated when $\kappa\geq 2.2$, then it crossovers to
 another rapid growth before it eventually saturates to a value 
 $\la\overline{\mathcal{F}}\ra_\mathcal{P}^\infty\approx0.18$ beyond $\kappa\approx5$.
 We notice here that the phase space averaged entanglement 
 as a function of chaoticity parameter $\kappa$ \cite{xwsg} behaves 
 very similar to $\la\overline{\mathcal{F}}\ra_\mathcal{P}$.
 
 Let us compare the behavior of
 $\la\overline{\mathcal{F}}\ra_\mathcal{P}$ with the classical LE to show how the signatures of the transition between regularity 
 and chaos are manifested themselves in $\la\overline{\mathcal{F}}\ra_\mathcal{P}$.
 It is known that for quantum systems that have well-defined classical limit,
 the quantum chaos in them is accompanied by classical chaos in its classical counterpart.
 Therefor, the results in Fig.~\ref{PFLY} imply that
 the onset of quantum chaos can be identified by the fast growth in 
 $\la\overline{\mathcal{F}}\ra_\mathcal{P}$ start from $\kappa\approx2.2$, 
 while the saturation of $\la\overline{\mathcal{F}}\ra_\mathcal{P}$ near 
 $\kappa\approx5$ indicates the appearing of the global chaos consistent with Ref.~\cite{xwsg}.

 On the other hand, Fig.~\ref{PFLY} also displays the obvious differences between 
 $\la\overline{\mathcal{F}}\ra_\mathcal{P}$ and classical LE.
 Unlike the classical LE, which equals zero in the regular region until $\kappa\approx2$,
 $\la\overline{\mathcal{F}}\ra_\mathcal{P}$ exhibits a rapid growth from a small 
 but nonzero value as $\kappa$ changes from $0$ to $1.8$.
 This increase of $\la\overline{\mathcal{F}}\ra_\mathcal{P}$ is due to the spread of the initial
 in phase space with increasing $\kappa$.
 Moreover, the saturation behavior of $\la\overline{\mathcal{F}}\ra_\mathcal{P}$ for finite quantum systems is 
 in contrast to the classical LE, which is still grows with $\kappa$ when $\kappa\geq5$.
 The saturation value $\la\overline{\mathcal{F}}\ra_\mathcal{P}^\infty$ increases with increasing system size $N$ and 
 one can expect that the plateau in $\la\overline{\mathcal{F}}\ra$ will disappears in the classical limit. 
 We also note as a consequence of the above mentioned slow polynomial growth of $\mathcal{F}(t)$ at the boundary between
 the regular and chaotic regions, the well-defined regularity-chaos transition located around $\kappa=2$ in classical system is
 replaced by the slow growth region in $\la\overline{\mathcal{F}}\ra_\mathcal{P}$ for quantum system. 
 In spite of these differences, we expect that a better quantum-classical correspondence can be established between
 the classical LE and $\la\overline{\mathcal{F}}\ra_\mathcal{P}$ in the classical limit ($N\to\infty$).   
 Therefore, our results confirm that the concept of bipartite fluctuations can serve
 as a good detector of quantum chaos.

 \section{Conclusion} \label{summary}

 To summarize, By using the kicked two-site BH model,
 we have shown that the concept of bipartite fluctuations is a useful diagnostic tool 
 to study the signatures of quantum chaos.
 We first demonstrated that the time evolution of bipartite fluctuations shows very different behaviors when 
 the location of initial coherent states changes from the regular to chaotic regions in classical phase space.
 Hence, the dynamics of bipartite fluctuations can uncover the signatures of quantum chaos.  
 More interestingly, we found that at the boundary between the regular and chaotic regions in classical phase space, the evolution
 of bipartite fluctuations exhibits a slow polynomial growth.
 The polynomial growth in the time evolution of local observables has been studied in several works for quantum systems that 
 lack of the well-defined classical counterpart \cite{Fine2014,Ivan2017,Silvia2019}.  
 By considering the quantum kicked BH model has a well-defined classical limit,
 we believe that understanding the mechanism of the 
 non-perturbative polynomial regime in quantum systems remains an open issue.
 
 We showed that the long-time averaged bipartite fluctuations provides a 
 measure of quantum chaos for each point in the classical phase space. 
 In regular regions it takes small values and approaching zero as the system size goes to infinity.
 In chaotic regions, it has finite values and independent of the system size.
 We further find the dynamics of bipartite fluctuations are intimately correlated to
 the localization properties of the evolved coherent state. 
 We finally discussed how to identify the global signatures of quantum chaos 
 from the behaviors of bipartite fluctuations power as a function of the chaoticity parameter,
 and compare it to the classical Lyapunov exponent.
 Even though bipartite fluctuations power has some differences from classical Lyapunov exponent,
 there is no doubt that bipartite fluctuations power allows us to clearly
 identify the characters of the transition between regularity and chaos.  
 Moreover, we can expect a much better correspondence between bipartite fluctuations power 
 and classical Lyapunov exponent in the classical limit.

 In the future it would be of interest to explore an analytical theory for our numerical results.
 In particular, connect bipartite fluctuations power to classical Lyapunov exponent in an analytical way
 could help us to advance our understanding of the quantum chaos.  
 Furthermore, as is pointed out in Refs.~\cite{Santos2004,Carlos2005}, 
 the relationship between quantum chaos and delocalization is a very delicate issue.
 To further confirm that the concept of bipartite fluctuations is a useful tool to study quantum chaos, 
 it could be beneficial to investigate bipartite fluctuations in other quantum chaotic systems,
 in particular, in quantum many-body chaotic systems. 
 Another interesting problem for future work is to characterize chaos in the quantum energy spectrum using 
 the concept of bipartite fluctuations.

 We stress that our numerical results for bipartite fluctuations clearly confirm the usefulness of the concept of bipartite 
 fluctuations in the studies of quantum chaos.
 By considering the two coupled tightly Bose-Einstein condensates has been proposed as a potential realization of the kicked
 two-site BH model \cite{marg,gross,riedel} and the experimental measurement of bipartite fluctuations
 can be achieved using single atom microscopy \cite{bakr,sherson}.
 We hope that our work would lead to more studies, both theoretical and experimental, on the
 signatures of quantum chaos using the concept of bipartite fluctuations.

 \section*{Acknowledgements}

 Q.W. is very grateful to Meng Sun for many useful discussions
 and hospitality at the Center for Theoretical Physics of Complex Systems,
 Institute for Basic Science, South Korea.
 This work has been supported by National Natural Science Foundation of China under grant No.~11805165
 and the Slovenian Research Agency (ARRS) under the grant J1-9112.

\end{document}